\documentclass[manuscript]{aastex}

\shorttitle{{\it RHESSI} Microflare Densities \& Filling Factors} 
\shortauthors{Baylor et al.}

\usepackage{graphicx}
\usepackage{dcolumn}
\usepackage{bm}
\usepackage{longtable}%
\usepackage{verbatim}

\begin{document}

\title{Estimates of Densities and Filling Factors from a Cooling Time
  Analysis of Solar Microflares Observed with {\it RHESSI}}

\author{R.~N.~Baylor\altaffilmark{1}, P.~A.~Cassak\altaffilmark{1},
  S.~Christe\altaffilmark{2}, I.~G.~Hannah\altaffilmark{3}, S\"{a}m
  Krucker\altaffilmark{4,5}, \\ D.~J.~Mullan\altaffilmark{6},
  M.~A.~Shay\altaffilmark{6}, H.~S.~Hudson\altaffilmark{3,4}, and
  R.~P.~Lin\altaffilmark{4,7}}

\altaffiltext{1}{Department of Physics, West Virginia University,
Morgantown, WV 26506, USA; Paul.Cassak@mail.wvu.edu}

\altaffiltext{2}{NASA Goddard Space Flight Center, Greenbelt, MD
20771, USA}

\altaffiltext{3}{School of Physics and Astronomy, University of
  Glasgow, Glasgow, Scotland, UK, G12 8QQ}

\altaffiltext{4}{Space Sciences Laboratory, University of California,
Berkeley, CA 94720-7450, USA}

\altaffiltext{5}{Institute of 4D Technologies, School of Engineering,
  University of Applied Sciences North Western Switzerland, 5210
  Windisch, Switzerland}

\altaffiltext{6}{Department of Physics and Astronomy and Bartol
Research Institute, University of Delaware, 217 Sharp Laboratory,
Newark, DE 19716, USA}

\altaffiltext{7}{Physics Department, University of California,
  Berkeley, CA 94720-7450 and School of Space Research, Kyung Hee
  University, Korea}

\begin{abstract}
  We use more than 4,500 microflares from the Reuven Ramaty High
  Energy Solar Spectroscopic Imager ({\it RHESSI}) microflare data set
  (Christe et al., 2008, Ap.~J., 677, 1385) to estimate electron
  densities and volumetric filling factors of microflare loops using a
  cooling time analysis.  We show that if the filling factor is
  assumed to be unity, the calculated conductive cooling times are
  much shorter than the observed flare decay times, which in turn are
  much shorter than the calculated radiative cooling times.  This is
  likely unphysical, but the contradiction can be resolved by assuming
  the radiative and conductive cooling times are comparable, which is
  valid when the flare loop temperature is a maximum and when external
  heating can be ignored.  We find that resultant radiative and
  conductive cooling times are comparable to observed decay times,
  which has been used as an assumption in some previous studies.  The
  inferred electron densities have a mean value of $10^{11.6} \ {\rm
    cm}^{-3}$ and filling factors have a mean of $10^{-3.7}$.  The
  filling factors are lower and densities are higher than previous
  estimates for large flares, but are similar to those found for two
  microflares by Moore et al.~(Ap.~J., 526, 505, 1999).
\end{abstract}




\maketitle

\section{INTRODUCTION}

Energy release in flares in solar active regions occurs over many
orders of magnitude, from large flares to microflares with as low as a
millionth the energy content of large flares.  The latter are A and B
{\it GOES} class events, which occur more frequently than large flares
with a negative power law distribution in number of flares as a
function of energy release extending over many decades in energy
\citep{Lin84,Dennis85,Crosby93,Feldman97,Wheatland00,Nita02,
  Paczuski05}, which suggests a common energy release mechanism.
There are some features in common among flares of all sizes, such as
radiation in multiple wavelength bands and similar X-ray light curves
[see, {\it e.g.}, \citet{Fletcher10} for a review].  However,
observational differences also exist between small and large flares.
Large flares are associated with higher temperatures than small flares
\citep{Feldman95,Caspi10}.  Also, weaker hard X-ray flares may have
steeper spectra than more energetic ones \citep{Christe08,Hannah08}.

Statistical studies of hard X-ray microflares have become more
comprehensive since the launch of the Reuven Ramaty High Energy Solar
Spectroscopic Imager ({\it RHESSI}) satellite \citep{Lin02}.  {\it
  RHESSI} achieves a lower energy cutoff in the X-ray spectrum than
previous detectors.  By using removable shutters, {\it RHESSI} allows
observation of both large and small flares [see, {\it e.g.},
\citet{Hannah10} for a description].  A recent study \citep{Christe08}
used a new flare-finding technique to identify over 24,000 microflares
from March 2002 - March 2007.  Statistical analyses of the microflare
properties were carried out by \citet{Christe08} and \citet{Hannah08}.
This large data set allows for unprecedented studies of flare
properties.

Here, we use this large data set to infer the electron density and the
volumetric filling factor of the microflare loops in the {\it RHESSI}
data set.  The volumetric filling factor is the fraction of the flare
loop volume from which radiation is detected.  While the filling
factor is often thought of as a robust parameter for flare loops, it
should be noted that its determination is potentially instrument- and
resolution-dependent.  A previous estimate of the density assumed the
filling factor was unity \citep{Hannah08}.  We use a cooling-time
analysis [see, {\it e.g.}, \citet{Moore80}], to argue that the
volumetric filling factors of microflare loops may be considerably
smaller than unity, implying densities considerably higher than the
estimates from \citet{Hannah08}.  The filling factors and densities we
find are consistent with a previous study of two microflares observed
with {\it Yohkoh} \citep{Moore99}.

The {\it RHESSI} microflare data set is described in
Sec.~\ref{sec-observations}.  The analytical technique is reviewed and
critiqued in Sec.~\ref{sec-technique}.  Results and uncertainty
estimates are presented in Sec.~\ref{sec-results}.  Finally,
conclusions are discussed in Sec.~\ref{sec-disc}.

\section{OBSERVATIONAL DATA}
\label{sec-observations}

A thorough description of the observational data is given by
\citet{Christe08} and \citet{Hannah08}; the details most salient for
the present study are summarized here.  The data set consists of all
microflares observed with {\it RHESSI} between March 2002 and March
2007.  The microflares were found to exclusively occur in active
regions.  The events were identified as local maxima in the count rate
of 6-12 keV photons having the appropriate sign of the time rate of
change of the count rate on either side of the maxima with
signal-to-noise ratio sufficiently large.  A total of 24,097 events
were identified.  Of these, spectral fitting and imaging analysis was
possible for 4,567 events, which allowed for a determination of the
plasma parameters required for the present analysis, namely the
microflare decay time $\tau_{D}$, the emission measure $EM$, the
temperature $T$, and the loop length $L$ and width $w$.

Histograms showing distributions of the parameters used in the present
study are given in Fig.~\ref{fig-rawdata}.  These and all
distributions use 45 equally sized bins.  The decay time $\tau_{D}$ of
the flare is determined from the flare finding algorithm
\citep{Christe08}, which employs a condition on the time derivative of
the count rate to determine the end of the event.  Their distribution
is shown in panel (a); the median value is 192 sec.  The minimum decay
time of any microflare in the present study is 4 sec.

The emission measure $EM$ (assumed isothermal) and temperature $T$ are
determined from spectral fitting of {\it RHESSI} hard X-ray data to a
model assuming an isothermal plasma with a power law tail above 10 keV
\citep{Hannah08}.  The logarithm of $EM$ is plotted in (b), with the
temperature in (c).  Median values of 13.0 MK and $2.38 \times
10^{46}$ cm$^{-3}$.  The uncertainty in $EM$ was estimated as $1 - 10
\%$ and the estimated (statistical) uncertainty in $T$ is less than $1
\%$ \citep{Hannah08}.

Finally, the physical size of the flare loops was estimated using
visibility forward fitting, described by \citet{Hannah08}, which fits
several Gaussian sources along the curved loop to estimate the
full-length $L$ and a central Gaussian FWHM to provide a measure of
the full-width $w$ of the flare loops.  The volume is estimated as $V
= \pi (w / 2)^{2} L$.  Distributions of $L, w,$ and the logarithm of
$V$ are plotted in panels (d), (e), and (f), respectively, with median
values of $2.09 \times 10^{9}$ cm, $0.665 \times 10^{9}$ cm, and $7.38
\times 10^{26}$ cm$^{3}$.  The estimated (statistical) uncertainty in
$L$ and $w$ is $\simeq 20 \%$, which is the standard deviation of
repeated (100) fit attempts with the visibility amplitudes randomized
within their statistical error each time \citep{Hannah08}.  There are
also systematic errors such as projection effects and the assumption
of a circular cross-section of the loops which are not included in the
estimate.  Another possible source of systematic error is that the
observations are from particles which have considerably higher energy
than the thermal background, so the determination of $L$ may be an
underestimate of the overall size of the loop because of the absence
of thermal particles in the data.

\section{DATA ANALYSIS}
\label{sec-technique}

The loop electron density $n_{e}$ was not measured, but can be
estimated using the definition of the isothermal emission measure
$EM$,
\begin{equation}
  EM = \int_{V} n_{e}^{2} dV,  \label{emdef}
\end{equation}
where $dV$ is a differential volume element and $V$ is the total
volume of a flare loop.  The simplest and most common way to estimate
$n_{e}$ is to assume it is uniform over the volume, which implies
\begin{equation}
  n_{e} = \sqrt{\frac{EM}{V}}. \label{denraw}
\end{equation}
This expression is correct if radiation can be detected from all
electrons which are present in the loop, {\it i.e.}, the loop is
assumed to be optically thin.  If finite optical depth effects are
present in a particular loop, then the observed X-ray flux from that
loop does not include direct contributions from all electrons, so the
actual density would be higher than the estimate in
Eq.~(\ref{denraw}).  Therefore, Eq.~(\ref{denraw}) provides a lower
bound on $n_{e}$.

An improvement on this technique comes from defining the so-called
filling factor $\phi$.  In terms of the filling factor, the
characteristic loop electron density $n_{e}$ is
\begin{equation}
  n_{e} = \sqrt{\frac{EM}{\phi V}}. \label{denfillfact}
\end{equation}
Since Eq.~(\ref{denraw}) gives a lower bound on $n_{e}$, $\phi$ is a
positive number between 0 and 1.

Here, estimates of the density will be tested using a cooling time
analysis.  Similar analyses have been performed previously in many
contexts \citep{Moore80,Haisch83,Stern83,Lin92,Cargill93,Moore99,
  Shibata99,Aschwanden00,Cargill04,Mullan06,Jiang06,Vrsnak06,
  Tsiropoula07,Cassak08a,Aschwanden08}.  Cooling time scales are
estimated using a scaling analysis (replacing derivatives by finite
differences of characteristic scales) of the hydrodynamic temperature
equation for a compressible optically-thin plasma,
\begin{equation}
  \frac{n k_{B}}{\gamma - 1} \frac{dT}{dt} = - n k_{B} T \nabla \cdot 
  {\bf v} + \kappa \nabla^{2} T - n_{e}^{2} \Lambda(T) + \dot{Q}_{ext}, 
  \label{tempeq}
\end{equation}
where $T$ is the temperature, $n$ is the {\it total} plasma density,
$\gamma$ is the ratio of specific heats, ${\bf v}$ is the bulk flow
velocity, $k_{B}$ is Boltzmann's constant, $\kappa(T)$ is the
coefficient of thermal conductivity, $\Lambda(T)$ is the radiative
loss function for an optically thin plasma, and $\dot{Q}_{ext}$ is the
volumetric heating rate from external sources.  Here, we assume
quasi-neutrality so that $n \simeq 2 n_{e}$ and that the plasma is an
ideal gas with $\gamma = 5 / 3$.  Comparing the left hand side to the
radiative loss term gives a radiative decay time scale $\tau_{R}$ of
\begin{equation}
  \tau_{R} \sim \frac{3k_{B}T}{n_{e} \Lambda(T)} \label{Rad}.
\end{equation}
Comparing the left hand side to the conduction term gives a conductive
decay time $\tau_{C}$ of
\begin{equation}
  \tau_{C} \sim \frac{3 n_{e} k_{B} (L/2)^{2}}{\kappa(T)}, \label{Cond}
\end{equation}
where $L/2$ is half the length of the flare loop which is the distance
from loop top to the solar surface.

The radiative loss function is usually taken as a piecewise continuous
function controlled by different physics at different temperatures.
For the temperatures of the flare plasmas in the present study [$T
\sim$ 8-20 MK from Fig.~\ref{fig-rawdata}(c)], the functional form of
the radiative loss function is $\Lambda(T) \simeq 5.49 \times
10^{-16}/T$ \citep{Klimchuk08}.  For the thermal conductivity
$\kappa(T)$, we employ the temperature-dependent parallel Spitzer
thermal conductivity of $\kappa(T) = \kappa_{0} T^{5/2} / \ln \lambda$
\citep{Spitzer53}, where the coefficient $\kappa_{0} = 1.84 \times
10^{-5} {\rm \ erg \ cm}^{-1} {\rm \ s}^{-1} {\rm \ K}^{-7/2}$, the
temperature is in Kelvin, and $\ln \lambda$ is the Coulomb logarithm
with $\lambda = (3 / 2e^{3})(k_{B}^{3}T^{3} / \pi n_{e})^{1/2}$ for a
pure hydrogen plasma.

This type of scaling analysis has been used previously to estimate
cooling times of flare loops and determine which mechanism dominates
the cooling, typically for large flares.  Early studies
\citep{Antiochos76} suggested conductive cooling is more efficient,
but the effects of chromospheric evaporation slow it down
\citep{Antiochos78}.  The role of radiation was studied
\citep{Antiochos80}, and for a while it was believed that radiation
and conduction act comparably to cool flare loops \citep{Moore80}
because the predicted times from the scaling analysis were comparable
to observed flare loop decay times $\tau_{D}$.  From the theoretical
perspective, it is reasonable that these time scales are comparable
due to the function of the chromosphere as a reservoir for the corona
\citep{Sturrock80,Moore80}.

The observational and theoretical result that $\tau_{C} \sim \tau_{R}
\sim \tau_{D}$ prompted authors to assume this relation to estimate
parameters for stellar flares for which well-resolved optical data
were not available \citep{Haisch83,Stern83}.  A recent study comparing
predictions using this model to independently derived parameters
($n,L,T$) of stellar loops found good agreement \citep{Mullan06},
lending credence to the validity of this assumption.

However, caution must be used in interpreting the scaling analysis
time scales as genuinely representative of the decay of flare loops.
\citet{Doschek82} used simulations to suggest that conduction
dominates early in time when the temperature is highest, followed by
comparable contributions from radiation and conduction.
\citet{Cargill93} used a model in which strictly conductive cooling
occurred at early times before transitioning to radiative cooling at
flare maximum because of chromospheric evaporation enhancing radiative
cooling at late times.  Therefore, there need not be a single dominant
mechanism throughout the duration of the event.

Another important issue is that the parameters that go into
Eqs.~(\ref{Rad}) and (\ref{Cond}) are tacitly assumed to be constant
and uniform, but the temperature changes as the loop cools and thus
the scaling results are not applicable to finding the time it takes to
cool from one temperature to another \citep{Cargill95}.  Taking into
account the change in temperature would require time integration
\citep{Culhane70, Svestka87,Aschwanden09}.  Since the difference
between actual cooling times and the scaling result can be
significant, the scaling analysis time scales only indicate
instantaneous time scales of cooling \citep{Cargill95}.  

The scaling analysis is on firmer theoretical footing at peak flare
temperature.  At peak temperature, the loop goes from being heated to
cooling, so the left hand side of Eq.~(\ref{tempeq}) is zero
instantaneously \citep{Aschwanden07}.  Ignoring the expansion term
(which is safe when flow speeds are subsonic) and assuming that there
is no external heating, the right-hand side implies that conduction
balances radiation instantaneously at the temperature peak, {\it
  i.e.,}
\begin{equation}
\tau_{C} \sim \tau_{R}. \label{tauequal}
\end{equation}
This is the same relation as before, but with a very different
interpretation.  For the purposes of the present study, we subscribe
to the latter interpretation of expecting equality only at peak flare
temperature rather than interpreting the scaling results as
predictions for the actual decay time.  (However, a relation to the
decay time will be discussed in the following section.)  Thus, the
cooling times are evaluated at the beginning of the cooling process.

It is important to point out aspects left out of the present model
that have been discussed in previous studies.  The scaling analysis
ignores pressure variation along the tube \citep{serio81}, radiative
cooling at loop footpoints \citep{Antiochos82}, chromospheric
evaporation \citep{Cargill95}, shrinkage of loops \citep{Svestka87b,
  Forbes96}, spatial nonuniformity \citep{Antiochos00}, and the effect
of multiple loops \citep{Reeves02}.  See \citet{Aschwanden08b} for an
approach incorporating fractal dimensional filling of flare loops.  A
recent study emphasizes the role of enthalpy flux in flare loops
\citep{Bradshaw10}.  If any of these aspects of solar flare evolution
play a significant role in determining the time-scales of flare decay,
the results of the present paper could change in detail.

\section{RESULTS}
\label{sec-results}

\begin{figure}
\epsscale{0.5}
\plotone{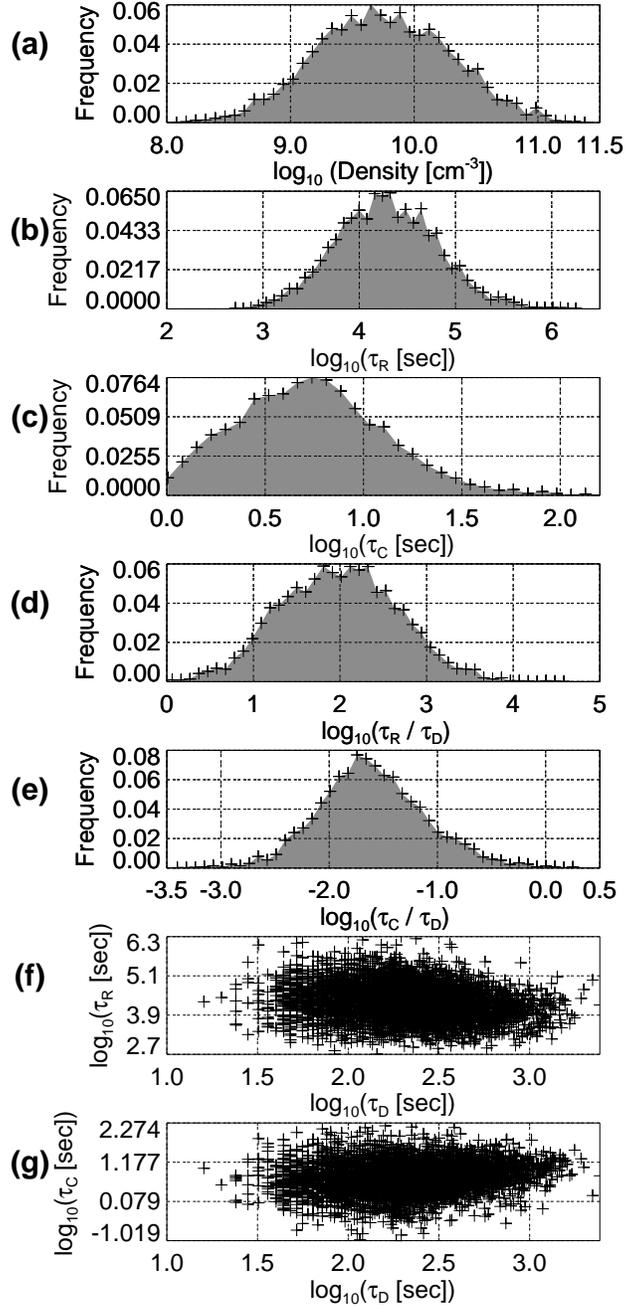}
\caption{\label{fig-ffequals1} Flare parameters assuming a filling
  factor $\phi$ of unity.  Plotted are distributions of the logarithms
  of (a) electron density $n_{e}$, (b) calculated values of the
  resistive cooling time $\tau_{R}$ and (c) the conductive cooling
  time $\tau_{C}$, (d) the ratio of radiative to flare decay times
  $\tau_{R} / \tau_{D}$, and (e) the ratio of conductive cooling times
  to flare decay times $\tau_{C} / \tau_{D}$.  Panels (f) and (g) are
  scatter plots of the logarithms of $\tau_{R}$ vs.~$\tau_{D}$ and
  $\tau_{C}$ vs.~$\tau_{D}$, respectively.}
\end{figure}

The density estimates presented in \citet{Hannah08} employed
Eq.~(\ref{denraw}) which assumes a filling factor of unity.  The
distribution of the logarithm of $n_{e}$ under this assumption for the
data in the present study is plotted in Fig.~\ref{fig-ffequals1}(a).
The mean electron density of the distribution in
Fig.~\ref{fig-ffequals1}(a) is $n_{e} \simeq 10^{9.8} \ {\rm
  cm}^{-3}$.  As noted earlier, this mean density is a lower bound on
the true mean density of the flares in the present study.

Using the density derived for each individual event, we estimate the
radiative and conductive cooling times using Eqs.~(\ref{Rad}) and
(\ref{Cond}) for each microflare.  The raw values for the
distributions of the logarithms of the resultant radiative and
conductive cooling times are plotted in Figs.~\ref{fig-ffequals1}(b)
and (c), respectively.  The median values of the calculated $\tau_{R}$
and $\tau_{C}$ are $2.06 \times 10^{4}$ sec and 5.43 sec,
respectively.  Figs.~\ref{fig-ffequals1}(d) and (e) show the same
values normalized to the flare's observed decay times $\tau_{D}$.
Panels (f) and (g) show a scatter plot of the logarithm of the
calculated cooling times compared to the logarithm of their decay
times, showing that the time scales are essentially uncorrelated.
Panel (d) shows that the radiative cooling times are distributed
around a peak about 100 times longer than $\tau_{D}$, while panel (e)
shows that the conductive cooling times are distributed around a peak
almost 100 times smaller than $\tau_{D}$, {\it i.e.,} $\tau_{C}\ll
\tau_{D} \ll \tau_{R}$.  This strongly contradicts the hypothesis in
Eq.~(\ref{tauequal}).  Note, this assessment assumes that it is
reasonable to compare the measured decay time to instantaneous
e-folding times from the scaling.  Since they may differ, this could
introduce systematic errors in the comparison.  However, we suspect
that it is not enough to account for the large separation in scales
inferred here.

Assuming the plasma parameters we use (including the density) are
correct, it is difficult to envision a physical explanation which
could justify the disparity in time scales because one expects a flare
to decay on a time-scale determined by the shortest available
dissipation time.  From this perspective, it is difficult to
understand how, in the presence of strong conductive cooling, the
actual decay time can be much longer than the conductive time-scale.
A likely cause is the assumption that $\phi = 1$, which we now relax.

In order to address the four orders of magnitude disparity which
appears to exist between the conductive and radiative decay times, we
assume Eq.~(\ref{tauequal}) holds and use it to solve for the density
and filling factor.  We then check to see if this assumption helps us
arrive at an internally consistent set of flare decay time scales.
Equating Eqs.~(\ref{Rad}) and (\ref{Cond}) and solving for $n_{e}$
gives
\begin{equation}
  n_{e} = \sqrt{\frac{4 \kappa(T) T}{L^{2}\Lambda(T)}}. \label{denff}
\end{equation}
This is equivalent to the classical analysis of \citet{Rosner78}.
Since $\kappa$ is a (weak) function of density due to the Coulomb
logarithm, we employ an iterative technique to self-consistently solve
for the density.  The procedure is to assume $\phi = 1$ to obtain a
zeroth order estimate $n_{0}$ which is used to calculate the zeroth
order $\kappa_{0}$.  The next order of density $n_{1}$ is then
determined from the previous $\kappa_{0}$.  This is continued until
convergence.  We find $n_{e}$ is determined to eight significant
figures after ten iterations and that the iterative procedure changes
$\Lambda(T)$ by only 10\%.  Using this value of the density, the
filling factor is obtained from Eq.~(\ref{denfillfact}) and cooling
times are obtained from Eqs.~(\ref{Rad}) and (\ref{Cond}).

\begin{figure}
\plotone{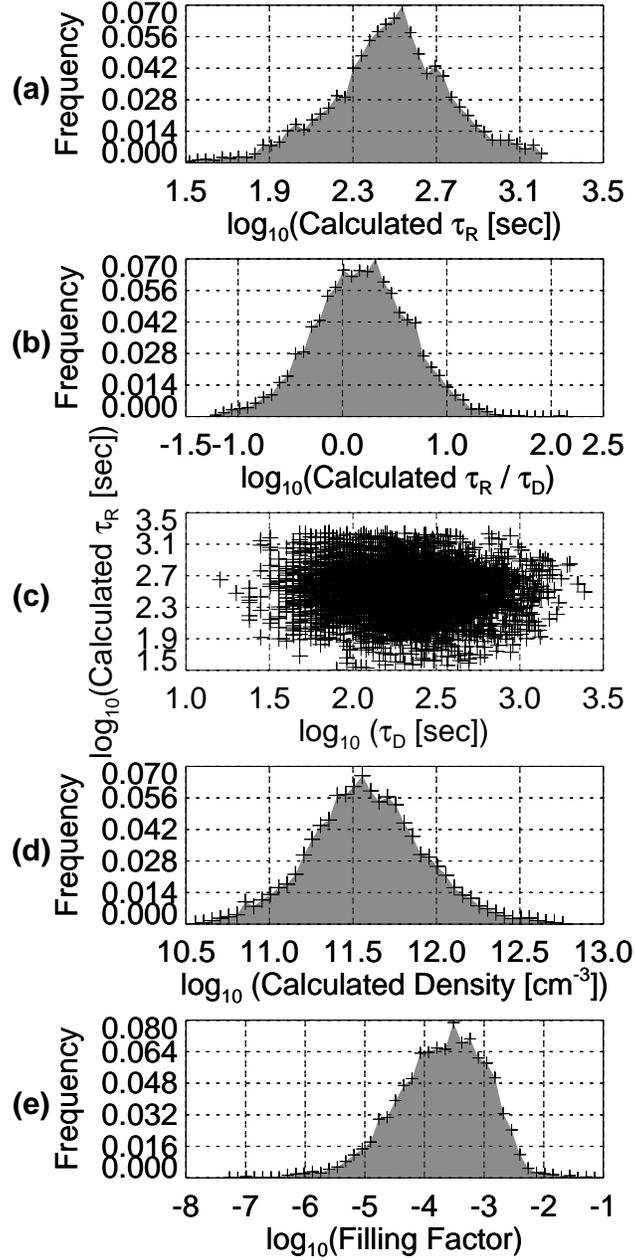}
\caption{\label{fig-ffresults} Flare parameters assuming radiative and
  conductive cooling times are equal.  Plotted are distributions of
  the logarithms of (a) the radiative cooling time $\tau_{R}$ and (b)
  the ratio of radiative to flare decay times $\tau_{R} / \tau_{D}$.
  The conductive cooling time is equivalent to the radiative cooling
  time by construction.  Panel (c) shows a scatter plot of the
  logarithms of $\tau_{R}$ vs.~$\tau_{D}$.  Also plotted are the
  logarithms of (d) the electron density $n_{e}$, and (e) the filling
  factor $\phi$.}
\end{figure}

The results of this analysis are displayed in
Fig.~\ref{fig-ffresults}.  Panel (a) shows the distribution of the
logarithm of the radiative cooling time $\tau_{R}$, while (b) shows
the distribution for the logarithm of $\tau_{R}$ normalized to the
flare decay time $\tau_{D}$.  The conductive cooling times $\tau_{C}$
are equal to $\tau_{R}$ by construction.  The median value of
$\tau_{R}$ is 325 sec, which is within a factor of 1.7 of the median
value of $\tau_{D}$.  This is reiterated in panel (c), which is a
scatter plot of $\tau_{R}$ and $\tau_{D}$.  The values do not appear
to be correlated, but they are clearly of the same order.  It is
surprising that the scales of the cooling times are essentially equal
to a key empirical time scale, $\tau_{D}$; nothing in the model
requires that such a similarity should emerge from the analysis.  Of
course, there are uncertainties associated with the comparison of
observed decay times and predicted e-folding scaling times, but the
present results lend observational support that a model with $\tau_{C}
\sim \tau_{R} \sim \tau_{D}$ (as has been often assumed before) is
consistent, at least for the present data set.

Panel (d) shows the logarithm of the resultant values for the
calculated densities from Eq.~(\ref{denff}).  The distribution has a
mean of $n_{e} \sim 10^{11.6} \ {\rm cm}^{-3}$, which is nearly two
orders of magnitude higher than the reported values in
\citet{Hannah08} and plotted in Fig.~\ref{fig-ffequals1}(a).  The
inferred filling factors, the logarithm of which is shown in panel
(e), have a mean of $\phi \sim 10^{-3.7}$.  We discuss these results
further in the following section.

\begin{figure}
\plotone{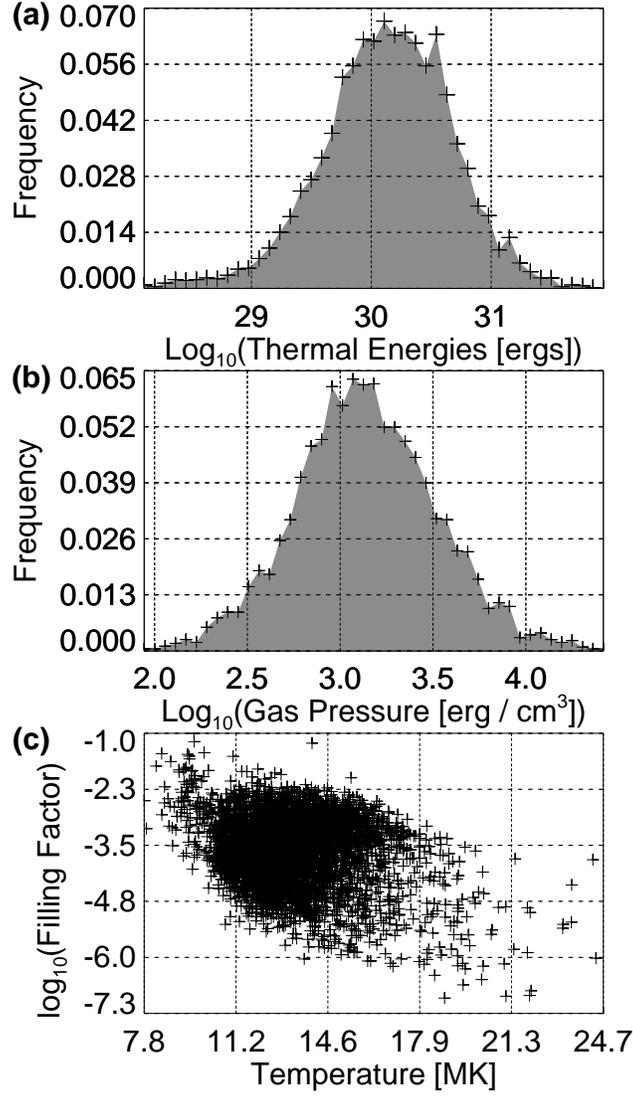}
\caption{\label{fig-calculated} Distributions of the logarithms of (a)
  the calculated thermal energy $W_{T} = 3 n_{e} k_{B} T_{e} V$ and
  (b) gas pressure $P = 2 n_{e} k_{B} T_{e}$ using the values of
  density from Fig.~\ref{fig-ffresults}(a).  (c) Scatter plot of the
  logarithm of the filling factor $\phi$ against electron temperature
  $T_{e}$.}
\end{figure}

Using the calculated densities, one can calculate other properties of
the flare loops.  The logarithm of the total thermal energy $W_{T} = 3
n_{e} k_{B} T_{e} V$ in the flare loops is shown in
Fig.~\ref{fig-calculated}(a), with a median value of $1.57 \times
10^{30}$ erg.  The distribution of the logarithm of the calculated gas
pressures $P = 2 n_{e} k_{B} T_{e}$ is shown in
Fig.~\ref{fig-calculated}(b), with a median value of $1.43 \times
10^{3}$ erg cm$^{-1}$.  Figure \ref{fig-calculated}(c) is a scatter
plot of the logarithm of the calculated filling factor $\phi$ against
the electron temperature $T_{e}$.  The filling factor is smaller for
higher temperature loops, which is a consequence of
Eqs.~(\ref{denfillfact}) and (\ref{denff}).

We now discuss the effect of uncertainties in the present analysis.
As discussed in Sec.~\ref{sec-observations}, the statistical errors in
the length $L$, emission measure $EM$, and temperature $T$ are
approximately $20\%, 10\%$, and $1\%$ \citep{Christe08,Hannah08}.
Standard error propagation techniques imply uncertainties for
calculated cooling times of $40\%$, densities $20\%$, and filling
factors $40\%$, which are sizable but not unreasonably large.

A few potential sources of systematic errors have been noted.  They
include assuming that the measured decay time $\tau_{D}$ corresponds
to an e-folding decay time from a scaling analysis, that the
representative loop length is assumed to be equal to the values
obtained by \citet{Hannah08} determined from particles at energies far
above thermal energies, and that there is no external heating at the
time of peak flare temperature.

\section{DISCUSSION}
\label{sec-disc}

In this paper, {\it RHESSI} microflare data are used to estimate the
volumetric filling factor and the electron density of microflare loops
using an analysis of cooling times.  If the filling factor is assumed
to be unity, then the conductive cooling time of the loop is much
smaller than the observed decay time, which itself is much smaller
than the radiative decay time.  This is difficult to justify
physically.  Alternately, if one invokes the hypothesis that the
radiative and conductive cooling times are comparable at the moment
when the flare temperature passes through its maximum value (and that
cooling due to expansion and flare heating are negligible at that
time), one can solve for the filling factor and density.  Mean values
for the whole distribution are $\phi \sim 10^{-3.7}$ and $n_{e} \sim
10^{11.6} \ {\rm cm}^{-3}$.  Our weakest assumption is that flare
heating stops at the peak time of hard X-rays.  Since the hard X-ray
time profile is a convolution of heating and cooling, heating does not
necessarily stop at the hard X-ray peak time.  If heating is present
during the decay, even at a low level, the cooling times could be
longer than derived for the case without heating, and the filling
factor could be larger than derived here.

Our estimate of mean densities are higher than those reported by
\citet{Hannah08}.  The authors are aware of only one other systematic
study of microflare loop densities or filling factors \citep{Moore99},
who used {\it Yohkoh} to study two microflare strands.  Using an
identical analytical technique as the present study, they found
densities between $n_{e} \sim 10^{10} \ {\rm cm}^{-3}$ and $10^{11.6}
\ {\rm cm}^{-3}$, with filling factors between $10^{-3.2}$ and
$10^{-2.8}$.  Thus, the values in the two microflares studied by
\citep{Moore99} from {\it Yohkoh} are in good agreement with results
from the large {\it RHESSI} data set studied here.

We now compare the present results with observations of large flares.
\citet{Culhane94} found $\phi \sim 1$ and $n_{e} \sim 3 \times 10^{11}
\ {\rm cm}^{-3}$ for an M-class flare, \citet{Varady00} found $\phi
\sim 0.01-0.2$ and $n_{e} \sim 7 \times 10^{9} - 1.5 \times 10^{10} \
{\rm cm}^{-3}$ for a C-class flare, \citet{Aschwanden01b} found $\phi
\sim 1$ and $n_{e} \sim 1.5 \times 10^{10} \ {\rm cm}^{-3}$ for the
Bastille Day flare, \citet{Teriaca06} found $\phi \sim 0.2-0.5$ and
$n_{e} \sim 10^{10} \ {\rm cm}^{-3}$ for a C-class flare, and
\citet{Raymond07} found $\phi \geq 0.01$ and $n_{e} \sim 10^{11} \
{\rm cm}^{-3}$ for X-class flares.  Other examples of filling factors
include 0.3 in active region loops \citep{Landi09}, 0.04-0.07 in
coronal holes \citep{Abramenko09}, and near unity in many coronal hole
jets \citep{Doschek10} but 0.03 in another study \citep{Chifor08}.  As
noted earlier, the determination of filling factors can depend on
detector resolution and wavelength.  Nonetheless, the filling factors
obtained here for microflares are at least 10 times smaller than those
reported for large flares, which is likely statistically significant.
Densities are slightly higher for microflares in the present study
than for larger flares in previous studies.

\begin{figure}
\plotone{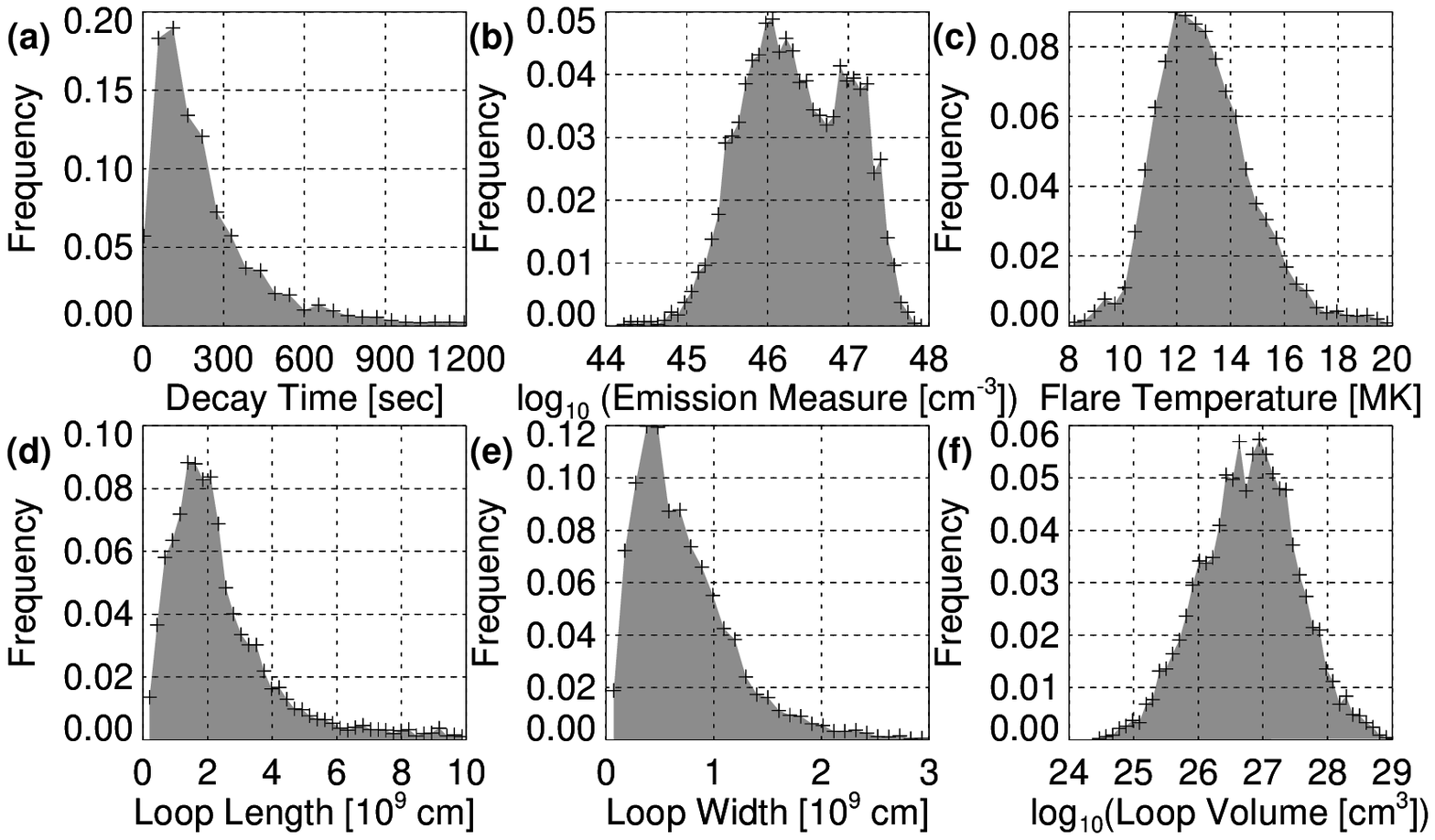}
\caption{\label{fig-rawdata} Distributions of raw data used in the
  present study using the methods described by \citet{Christe08} and
  \citet{Hannah08}.  Plotted are (a) flare decay time $\tau_{D}$, (b)
  the logarithm of the emission measure $EM$, (c) the logarithm of the
  temperature $T$, (d) flare loop full-length $L$, (e) flare loop
  full-width $w$, and (f) the logarithm of the flare loop volume $V$.}
\end{figure}

Given the characteristics of the microflare data set considered here
compared to large flares, it is perhaps not surprising that the
filling factors are small.  The microflare loops in the present study
occur in active regions, just as large flares do.  The mean sizes of
the loops in the present study are comparable to those of larger
flares.  The loops have energies of $n_{e} k_{B} T V \sim 10^{28} {\rm
  \ ergs}$ deposited into them by the flare (using characteristic
values from Fig.~\ref{fig-rawdata}), which is about $10^{4}$ times
smaller than large flares.  Thus, the loops are of similar size but
acquire less energy, which could lead to a smaller filling factor.  To
estimate the size of the region for which radiation is detected, we
note the radiating volume is $V_{*} = \phi V$.  Assuming that the
length of the radiating plasma is $L$, the effective width $w_{*}$ of
the radiating plasma is given by $V_{*} \sim \pi (w_{*} / 2)^{2} L$.
For the present parameters, this implies loops with total thickness
$w_{*} \sim \phi^{1/2} w \simeq w / 100 \simeq 4 \times 10^{6} \ {\rm
  cm}$, which would imply there is an unseen substructure of thin
strands within the flare loops.

Various lines of evidence indicate that there are smaller-scale
structures in the corona, e.g.~\citet{Mullan90}.  Radio polarization
data point to the existence of structures in the corona which are
$\sim ~100$ km in size \citep{Melrose75}.  The possibility that 100 km
structures are associated with collapsing magnetic reconnection sites
in the corona was discussed by \citet{Mullan80}: using constraints on
the collapse time-scales and coronal Alfv\'en speeds, transverse
dimensions of order 100 km were found to be typical of reconnection
sites in the corona.  More recently, there is evidence that X-class
flare loops are composed of thin threads from high resolution
observations, with structure at scales of a few arcseconds ($1''
\simeq 10^{8} {\rm \ cm}$) and below \citep{Dennis09,Kontar10,
  Krucker10}.  There is also abundant evidence from the footpoints of
flaring loops that most of the emission is spatially unresolved, such
as in TRACE white-light flares \citep{Hudson06}.  \citet{Xu06} found a
core region within a halo region in two X-class white-light flares,
reporting a ratio of the area of the core to the halo of 4\% and 25\%,
respectively.  Also, simulations of loops comprised of many small
scale filaments were able to reproduce cooling characteristics of
large flare loops observed with TRACE \citep{Warren03}.  Thus, the
conclusion that there is small substructure of flare loops is not
without precedent.

The prediction of small scale loops has implications for the heating
mechanism of the flare loops.  \citet{Hannah08} estimated that the
non-thermal power in accelerated electrons during the time of peak
emission in the {\it RHESSI} microflares is $10^{26} \ {\rm erg \
  s}^{-1}$.  For loops of area $10^{14} {\rm \ cm}^{2}$ as predicted
by the present results, the energy deposition rate per unit area would
be $10^{12} {\rm \ erg \ s}^{-1} {\rm \ cm}^{-2}$.  This is an
enormous value, as discussed in \citet{Krucker10}.  Hence, if the
filling factor is indeed $\sim 10^{-4}$, then microflares are unlikely
heated by electron beams.  A recent model that the flare energy is
transported by Alfv\'en waves \citep{Fletcher08} would not be ruled
out by the data.

An interesting result of the present study is that the conductive and
radiative cooling times derived by assuming their equality at the time
of maximum temperature are comparable to the observed microflare decay
times.  A possible ramification of this result is that it lends
credence to the assumption that the conductive and radiative times are
comparable to the decay time, $\tau_{R} \sim \tau_{C} \sim \tau_{D}$.
This is relevant to stellar flare studies in which plasma parameters
were obtained under such assumptions \citep{Haisch83,Stern83}.  In
addition, a previous study of stellar flares \citep{Mullan06} found
results of this model are largely consistent with independently
determined plasma parameters.  If one believes that the scaling
analysis cooling times actually represent physical cooling times for
loops, the result suggests that conductive and radiative cooling act
at comparable levels to cool flare loops, at least for the microflares
in the present study.

Future work could include efforts to incorporate physical effects left
out of the model as summarized at the end of Sec.~\ref{sec-technique}.
Also, future studies could further try to minimize the systematic
errors discussed in Sec.~\ref{sec-results}.  These can be addressed
both with observations and with numerical modeling.  Also, the study
of the cooling times of individual events will help determine the
validity of the cooling time analysis.

The authors thank G.~Holman and B.~Dennis for helpful conversations.
RNB and PAC gratefully acknowledge support by NSF grant PHY-0902479,
NASA's EPSCoR Research Infrastructure Development Program and the West
Virginia University Faculty Senate Research Grant program.  DJM is
supported in part by the Delaware Space Grant.  IGH is supported by a
STFC rolling grant and by the European Commission through the SOLAIRE
Network (MTRN-CT-2006-035484).  RPL is supported in part by the WCU
grant (No. R31-10016) funded by the Korean Ministry of Education,
Science and Technology.  HSH, RPL, and SK are supported through NASA
contract NAS 5-98033 for {\it RHESSI}.



\clearpage

\clearpage

\clearpage

\clearpage

\end{document}